\documentclass[aps,reprint,pre]{revtex4-1}

\usepackage{graphicx}  
\usepackage{bm}        
\usepackage{amssymb,amsmath}
\usepackage{xcolor}
\usepackage{mathtools}

\newcommand{\one}{\mathbb I} 
\newcommand{\E}{\mathbb E} 

\newcommand{\diag}{\operatorname{diag}} 

\renewcommand{\phi}{\varphi} 
\renewcommand{\epsilon}{\varepsilon} 

\DeclarePairedDelimiter{\abs}{\lvert}{\rvert} 

\begin{document}

\title{Consequences of Dale's law on the stability-complexity relationship\\ of random neural networks }


\author{J. R. Ipsen}\email{jesper.ipsen@unimelb.edu.au}
\affiliation{ARC Centre of Excellence for Mathematical and Statistical Frontiers, School of Mathematics and Statistics, The University of Melbourne, 3010 Parkville, VIC, Australia}

\author{A. D. H. Peterson}\email{peterson@unimelb.edu.au}
\affiliation {Graeme Clarke Institute, The University of Melbourne, 3053 Carlton, VIC, Australia}
\affiliation {Department of Medicine, St. Vincent's Hospital, The University of Melbourne, 3065 Fitzroy, VIC, Australia}

\begin{abstract}
\noindent
In the study of randomly connected neural network dynamics there is a phase transition from a `simple' state with few equilibria to a `complex' state characterised by the number of equilibria growing exponentially with the neuron population.  Such phase transitions are often used to describe pathological brain state transitions observed in neurological diseases such as epilepsy.  In this paper we investigate how more realistic heterogeneous network structures affect these phase transitions using techniques from random matrix theory. Specifically, we parameterise the network structure according to Dale's Law and use the Kac-Rice formalism to compute the change in the number of equilibria when a phase transition occurs.  We also examine the condition where the network is not balanced between excitation and inhibition causing outliers to appear in the eigenspectrum.  This enables us to compute the effects of different heterogeneous network connectivities on brain state transitions, which can provide new insights into pathological brain dynamics.  
\end{abstract}

\date{\today}

\maketitle

\section{Introduction}

One of the major difficulties in the study of brain dynamics and neurological disease is that it is highly patient-specific and varies significantly between individuals.  One way of capturing these heterogeneous differences mathematically is to quantify different brain connectivities and examine their effects on brain network dynamics, particularly brain state transitions.  Typically, this directed graph or networked dynamical system is described using neural mass models where the network connectivity is either averaged over, losing its individuation or via high dimensional brute force numerical simulations, which are not mathematically tractable.  
An alternative approach is to study randomly connected neural networks using mean-field theory~\cite{Sompolinsky1988}, where it has been found that there is a rapid transition to a complex macroscopic `chaotic' state~\cite{Stern2014}.  Recently, in relation to this there have been two major mathematical developments: First, this transition has been described microscopically by Wainrib and Toboul~\cite{Wainrib2013} as an exponential explosion in the number of fixed points of the system.  Second, Rajan and Abbott~\cite{Rajan2006} discovered that more anatomically realistic non-random connectivity structures such as Dale's law change the uniform density of the eigenspectrum.  In this paper we combine these two seminal papers to show the effects of more anatomically realistic connectivity statistics on phase transitions.  

We extend and analyse a random neural network model as previously studied  using dynamical mean-field theory by Sompolinsky et al. \cite{Hansel1992, Sompolinsky1988}.  This is a first order neural model with a time-constant and discrete neural field described by a random connectivity matrix.  The model itself can be derived from a neural mass model by assuming instantaneous synapses.  

In such a system it was found that the dynamics sharply transitioned into a macroscopic chaotic regime with the variance of the connectivity matrix as the order parameter.  More recently, Wainrib and Touboul~\cite{Wainrib2013} have provided a microscopic explanation in terms of a transition from a system with a single equilibrium to a system with exponentially many equilibria.  However, both of these analyses have assumed an  i.i.d. (independent and identically distributed) random network connectivity that is biologically unrealistic.  
 
Anatomically, neurons connect via Dale's law \cite{Eccles1976}, which states that neurons that are excitatory (inhibitory)  only output excitatory (inhibitory) signals.  This gives the connectivity matrix a heterogeneous block structure where each block represents a neural population that draws from independent distributions with different means and variances.  In a seminal work by Rajan and Abbott \cite{Rajan2006}, it was shown that when these distributions have different variances, the eigenspectrum usually described by the so-called Circular Law~\cite{Bordenave2012} no longer has a uniform density.  In \cite{Ahmadian2015}, this structure was further generalised to a much larger class of connectivity matrices constructed as a combination of random and deterministic matrices incorporating non-random aspects of network connectivity.  A description of block structured matrices and how this affects the transition to chaos is also given by \cite{Aljadeff2015prl,Aljadeff2015jmp}. It has also been shown that there are effects when using a non-zero mean connectivity matrix, which introduces outliers to the eigenspectrum; this was originally pointed out by \cite{lang1964isolated} for real symmetric matrices.

This paper is organised as follows: in Section~\ref{sec:dale} we describe a mathematical framework of how to incorporate additional deterministic network structure into the otherwise random neural network.  Specifically, we incorporate Dale's law into the connectivity matrix. Importantly, we give a considerably simplified expression of the result originally presented by Rajan and Abbott~\cite{Rajan2006}. In Section~\ref{sec:complexity}, we combine the work of Wainrib and Touboul \cite{Wainrib2013} and Rajan and Abbott \cite{Rajan2006} to explore how Dale's law affects the microscopic description of the stability transition in terms of the number of fixed points. In Section~\ref{sec:outlier}, we discuss the effects of spectral outliers. The final section is devoted to a discussion of the results.

\section{Dale's law in neural models}
\label{sec:dale}

Consider the first order rate-based neural model for the evolution of a neural network given by~\cite{Sompolinsky1988}
\begin{equation}\label{dyn-sys}
\frac{dx_i}{dt}=-\frac{x_i}\tau+\sum_{j=1}^nJ_{ij}S(x_j),\qquad i=1,\ldots,n
\end{equation}
where $J_{ij}$ is the (random) synaptic connection weight from the $j$-th to the $i$-th neuron, $\tau$ is the membrane relaxation time, and $S$ is an odd sigmoid function with unit slope at the origin. More precisely, we assume that $S$ is a smooth and a (strictly) monotonically increasing function with $S(0)=0$, $S'(0)=1$ and, $S(x)\to\pm1$ for $x\to\pm\infty$.
 
Knowledge about the structure of the synaptic connectivity matrix $J=(J_{ij})$ is essential for studies of the evolution of the dynamical system~\eqref{dyn-sys}. In particular, the eigenvalue with the largest real part plays a crucial role in determining whether the system dynamics are stable or chaotic. 
The most frequently used assumption is to take the entries of the connectivity matrix $J=(J_{ij})$ to be i.i.d. random variables with zero mean, variance $\sigma^2$, and finite fourth moment (Gaussian random variables are often chosen for simplicity). 
Under these assumptions, the eigenvalues of the connectivity matrix are uniformly distributed within a disk of radius, $\sigma\sqrt n$~(see the review~\cite{Bordenave2012} and references within) and importantly there are no outliers in the asymptotic limit. Using mean field techniques, it can be shown that this model has a phase transition into a chaotic regime at $\sigma\tau=n^{-1/2}$ in the limit of large neuron populations ($n\to\infty$)~\cite{Sompolinsky1988}. However, the mean field approach does not provide an explanation of the microscopic origin of this phase transition. Recently, it has been argued by Wainrib and Touboul~\cite{Wainrib2013} that the phase transition originate from an explosion in the number of equilibria. More precisely, one phase is characterised by a scenario in which the dynamical system~\eqref{dyn-sys} has a single stable equilibrium point, while the other phase is characterised by a scenario in which the average number of equilibria increases exponentially with the neuron population $n$.
 
Unfortunately, the assumption of i.i.d. entries does not apply to more realistic neural networks, where known anatomical and physiological constraints must be imposed. 
Anatomically, there are broadly speaking two types of neurons that are either excitatory or inhibitory.  These have different connectivities depending on the spatial scale. Physiologically, they obey Dale's law \cite{Eccles1976}, that is that each neuron only sends out either excitatory or inhibitory signals depending on what type of neuron it is.  This means that each column (or row) in the connectivity matrix must have the same sign.
Rajan and Abbott~\cite{Rajan2006} studied the effect of Dale's law. They split the entries of the synaptic connectivity matrix into two distributions or blocks with different means and variances representing excitatory and inhibitory neurons, such that $\E\{J_{ij}\}>0$ for $i=1,\ldots,k$ (excitatory) and $\E\{J_{ij}\}<0$ for $i=k+1,\ldots,n$ (inhibitory) for some $0\leq k\leq n$.
Ahmadian et al.~\cite{Ahmadian2015} proposed an even more general connectivity matrix incorporating both random and deterministic aspects of the form
\begin{equation}\label{connection-structure}
J=LAR+M.
\end{equation}
Here $A$ is a random matrix with entries which are i.i.d. random variables with zero mean and unit variance, while $L$, $R$, and $M$ are deterministic matrices incorporating non-random aspects of the connectivity matrix. 

Let us consider how to incorporate the effects of Dale's law within the structure~\eqref{connection-structure}. First, let $0<f<1$ denote the fraction of excitatory neurons. Following~\cite{Rajan2006}, we want to represent connections for excitatory (inhibitory) neurons by random variables with mean $\mu_E>0$ ($\mu_I<0$) and variance $\sigma_E^2$ ($\sigma_I^2$). In order to construct this, we introduce a diagonal matrix,
\begin{equation}\label{Sigma}
\Sigma=\diag(\underbrace{\sigma_E,\ldots,\sigma_E}_{nf\text{ times}},\underbrace{\sigma_I,\ldots,\sigma_I}_{n(1-f)\text{ times}}),
\end{equation}
and vectors
\begin{equation}\label{uv}
u=(1,\ldots,1)^T,\quad\ 
v=(\underbrace{\mu_E,\ldots,\mu_E}_{nf\text{ times}},\underbrace{\mu_I,\ldots,\mu_I}_{n(1-f)\text{ times}})^T.
\end{equation}
We can now choose the network connectivity matrix~\eqref{connection-structure} with $L=\one$, $R=\Sigma$, and $M=uv^T$, i.e.
\begin{equation}\label{J-outlier}
J=A\Sigma+uv^T.
\end{equation}
It is easily verified that this choice of $L$, $R$, and $M$ implies that $J_{ij}$ has mean $\mu_E$ ($\mu_I$) and variance $\sigma_E^2$ ($\sigma_I^2$) for $j<nf$ ($j>nf$), thus creating the desired distinction between excitatory and inhibitory neurons and implementing Dale's law.  
 
There is strong experimental evidence that the brain is very tightly balanced on multiple scales  \cite{Dehghani2016,Barral2016}. In the above given framework, balanced is taken to mean that
\begin{equation}\label{balance}
\mu_Ef+\mu_I(1-f)=0,
\end{equation}
or in other words, the combined contribution of excitatory and inhibitory neurons sums to zero on average.
\begin{figure}[htbp]
\centering
\includegraphics[width=.99\columnwidth]{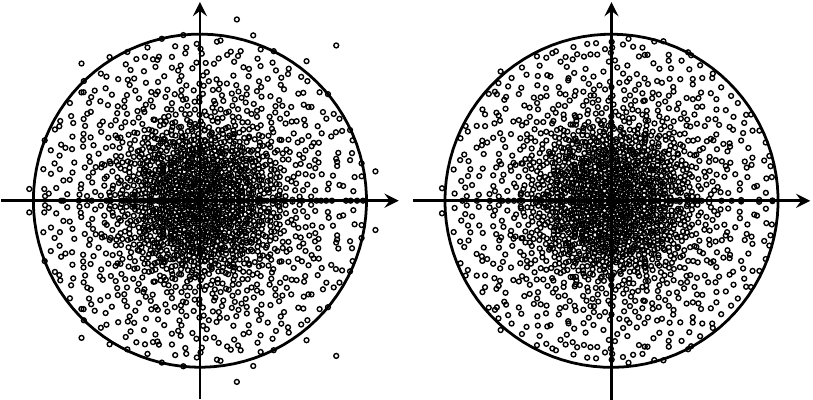}
\caption{The figure shows the scatter plots in the complex plane for eigenvalues of connectivity 
matrices~\eqref{J-outlier} and~\eqref{J-no_outlier} shown as the left and right panel, respectively. Both scatter plots are generated from random matrices with $n=2\,000$, $f=1/4$, $\mu_E=3$, $\mu_I=-1$, $\sigma_E=2$, and $\sigma_I=1/2$. Solid curves shown are circles with radius given by~\eqref{radius}.}
\label{fig:scatter}
\end{figure}
As illustrated by figure~\ref{fig:scatter} (left panel), after incorporating Dale's law, the vast majority of the eigenvalues of the connectivity matrix $J$ lies within a circular region with radius
\begin{equation}\label{radius}
R=\sigma_\text{eff}\sqrt n=\sqrt{n(f\sigma_E^2+(1-f)\sigma_I^2)}.
\end{equation}
Here, we have introduced an `effective' variance defined as $\sigma_\text{eff}^2=f\sigma_E^2+(1-f)\sigma_I^2$.
However, we also note that there are a number of outliers. These outliers do not disappear in the limit of large neuron populations $n\to\infty$. We will discuss the importance of such outliers in Section~\ref{sec:outlier}. The fact that imposing Dale's law introduces outliers to the eigenvalue spectrum of the connectivity matrix was first observed numerically in~\cite{Rajan2006}, while the existence of a unique limiting distribution for these outliers was proved rigorously in~\cite{tao2013outliers}.  
 
In order to remove the outliers, Rajan and Abbott~\cite{Rajan2006}  introduced an additional zero-sum constraint on the connectivity matrix, which may be written as
\begin{equation}\label{zero-sum}
\sum_{j=1}^n(J_{ij}-M_{ij})=0.
\end{equation}
Put together with the balance condition~\eqref{balance}, the zero-sum constraint~\eqref{zero-sum} states that the $i$-th neuron has a strict input-output balance. Here, `{strict}' is used to emphasise that the input-output balance holds not only on average but also in every realisation.

In order to incorporate the zero-sum constraint~\eqref{zero-sum}, we introduce the projection matrix
\begin{equation}
P=\one-\frac{uu^T}n
\end{equation}
with $u$ as in~\eqref{uv}. We consider a network connectivity matrix
\begin{equation}\label{J-no_outlier}
J=A\Sigma P+uv^T.
\end{equation}
This connectivity matrix closely resembles our previous choice~\eqref{J-outlier}, but the projection matrix, $P$, enforces the row zero-sum condition~\eqref{zero-sum}. Importantly, it can be shown that this new connectivity matrix $J$ has spectral radius~\eqref{radius} to leading order in $n$, hence there are no outliers (see~\cite{tao2013outliers} for a proof in the case with $\Sigma=\one$). Figure~\ref{fig:scatter} (right panel) provides a numerical verification of this phenomenon.
 
Due to the projection matrix $P$ in~\eqref{J-no_outlier}, the entries of $J$ are no longer independent. Nonetheless, it can still be shown that entries representing excitatory neurons have mean $\mu_E$ and variance $\sigma_E^2+O(n^{-1})$, while entries representing inhibitory neurons have mean $\mu_I$ and variance $\sigma_I^2+O(n^{-1})$. Thus, we still keep the desired distinction between excitatory and inhibitory neurons to leading order of the population size $n$.
 
The main difference between this network~\eqref{J-no_outlier} and the canonical example (in which the entries of $J$ are i.i.d. random variables) is that the eigenvalues are no longer uniformly distributed within the region of circular support. 
The global spectral density for the connectivity matrix~\eqref{J-no_outlier} was found by Rajan and Abbott~\cite{Rajan2006}, and we have considerably simplified their expression to
\begin{equation}\label{rho-RA}
\rho_\text{RA}(z)=
\begin{cases}
\displaystyle
\frac1{\pi n\sigma_I^2}\Big(1-\frac{g}{2}H_{f}\Big(g\frac{\abs{z}^2}{n\sigma_I^2}\Big)\Big),
& \abs z\leq \sigma_\text{eff}\sqrt n\\
0, & \abs z>\sigma_\text{eff}\sqrt n
\end{cases}
\end{equation}
with $g=1-\sigma_I^2/\sigma_E^2$  and
\begin{equation}\label{rho-RA-part2}
H_f(x)=\frac{2f-1+x+\sqrt{1+x(4f-2+x)}}{\sqrt{1+x(4f-2+x)}}.
\end{equation}
The derivation of the global spectral density~\eqref{rho-RA} assumes that $0<\sigma_I\leq\sigma_E$ and thereby $0\leq g< 1$, but by symmetry an equivalent formula holds for $\sigma_I\geq\sigma_E$ with $\sigma_I$ and $\sigma_E$ interchanged. Equivalency between our expression~\eqref{rho-RA} for the global spectral density and the expression provided by Rajan and Abbott in~\cite{Rajan2006} is easily verified using mathematical software such as \textsc{Mathematica}.

We note that for $\sigma_E=\sigma_I=\sigma$, we have $g=0$ and the eigenvalues are distributed uniformly within a disk with radius $\sigma\sqrt n$. Thus, from a spectral perspective this situation is equivalent to the case where $J$ has i.i.d. entries (changing to the rescaled variables $\hat z=z\sigma\sqrt n$ gives the usual situation with support on the unit disk). Likewise, the spectral density becomes uniform within a disk with radius $\sigma_E\sqrt n$ if $f\to0$ and uniform within a disk with radius $\sigma_I\sqrt n$ if $f\to1$.

\section{Counting equilibria as a measure of complexity}\label{sec:complexity}

In this section, we will be interested in the number of equilibria (both stable and unstable) for the dynamical system~\eqref{dyn-sys}, i.e. the number of solutions to
\begin{equation}\label{eq}
 0=-x_i/\tau+\sum_{j=1}^nJ_{ij}S(x_j),\qquad i=1,\ldots,n,
\end{equation}
with connectivity matrix $J=(J_{ij})$ given by~\eqref{J-no_outlier}.
It is clear that the number of equilibria is a random variable, since the connectivity matrix $J$ is a random matrix. We will denote this random variable by $N_\text{eq}$. 
Na\"ively, we may think of a system with few equilibria as ``simple'' and a system with many equilibria as ``complex''. In order to make this notion more concrete, we introduce the quantity
\begin{equation}\label{complexity-def}
C(\tau)=\lim_{n\to\infty}\frac1n\log\E\{N_\text{eq}\}\geq 0
\end{equation}
as a formal measure of complexity. 
The complexity measure~\eqref{complexity-def} --- henceforth referred to simply as \emph{complexity} --- plays a r\^ole similar to the free energy in statistical mechanics, and we will say that the system has a phase transition at $\tau_c$ if $C(\tau)$ is non-analytic at $\tau=\tau_c$.
We will consider the system as ``complex'' if $C(\tau)>0$ which implies that the mean number of equilibria grows exponentially fast with the neuron population $n$. If $C(\tau)=0$ then we consider the system as ``simple''. We will see that the neural network under consideration has a phase transition between a ``simple'' and ``complex'' phase and that this phase transition coincide with the transition predicted using mean field theory. 

The average number of equilibria can be obtained by means of the multivariate Kac--Rice formula (see e.g.~\cite{Adler2009,Azais2009,Fyodorov2015}),
\begin{multline}\label{KacRice}
\E\{N_\text{eq}\}=\int_{\mathbb R^n}d^nx\,
\E\bigg\{ \abs[\Big]{\det_{1\leq i,j\leq n}\Big[-\frac{\delta_{ij}}{\tau}+J_{ij}S'(x_j)\Big]}\\
\times\prod_{k=1}^n\delta\Big(-\frac{x_k}{\tau}+\sum_{\ell=1}^nJ_{k\ell}S(x_\ell)\Big)\bigg\}.
\end{multline}
For a few models the average number of fixed points may be calculated exactly, see e.g.~\cite{Ipsen2018}, but for the case considered here an approximation scheme is needed. We will borrow an approximation scheme suggested by Wainrib and Touboul~\cite{Wainrib2013}. We note that for short relaxation times (i.e. $\tau\to0$) the system~\eqref{dyn-sys} has a single (stable) equilibrium or, stated differently, equation~\eqref{eq} has a single solution. If the connectivity matrix $J$ has no eigenvalue outliers for a large neuron population $n\to\infty$ --- for our purposes that is the matrix \eqref{J-no_outlier} but not the matrix~\eqref{J-outlier} --- then it is assumed that the system~\eqref{dyn-sys} will have a single equilibrium up to the critical value $\tau_c$. Assuming that this is true, it was argued by Wainrib and Touboul~\cite{Wainrib2013} that the Kac--Rice formula~\eqref{KacRice} may be approximated by
\begin{align}
\E\{N_\text{eq}\}&\approx\E\bigg\{\abs[\Big]{\det_{1\leq i,j\leq n}\Big[-\delta_{ij}+\tau J_{ij}\Big]}\bigg\}
\label{KacRice-approx}
\end{align}
in the vicinity of the critical value $\tau\approx\tau_c$. 
%
It follows from~\eqref{complexity-def}, that the complexity becomes
\begin{equation}
C(\tau)
\approx\lim_{n\to\infty}\frac1n\log\E\bigg\{\abs[\Big]{\det_{1\leq i,j\leq n}[-\delta_{ij}+\tau J_{ij}]}\bigg\}.
\label{complexity-matrix}
\end{equation}
The right-hand side in~\eqref{complexity-matrix} only depends on the eigenvalues of the connectivity matrix $J$ and the parameter $\tau$; it is related to a what is known as a `linear statistic' in the random matrix literature. This quantity is expected to be self-averaging and we may approximate the complexity with a spectral integral 
\begin{equation}\label{complexity-general}
C(\tau)\approx\int_{\mathbb C} \mu_J(d^2z)\log\abs{\tau z-1},
\end{equation}
where $\mu_J(d^2z)$ denotes the global spectral measure (i.e. the limiting eigenvalue distribution) of the random connectivity matrix $J$ in the limit of large populations $n\gg1$.

So far our approximation has closely followed  the analysis performed by Wainrib and Touboul~\cite{Wainrib2013}. However, rather than considering an i.i.d connectivity matrix with a uniform spectral density, we consider a more structured matrix incorporating Dale's law that has a non-uniform spectral density~\eqref{rho-RA}, i.e. 
\begin{equation}
\mu_J(d^2z)=d^2z\rho_\text{RA}(z).
\end{equation}
As discussed this corresponds to a more anatomically realistic connectivity matrix where the connectivity for different populations are drawn from different distributions resulting in a partially random matrix that is block structured.  
 
\begin{figure}
 \centering
 \includegraphics[width=.99\columnwidth]{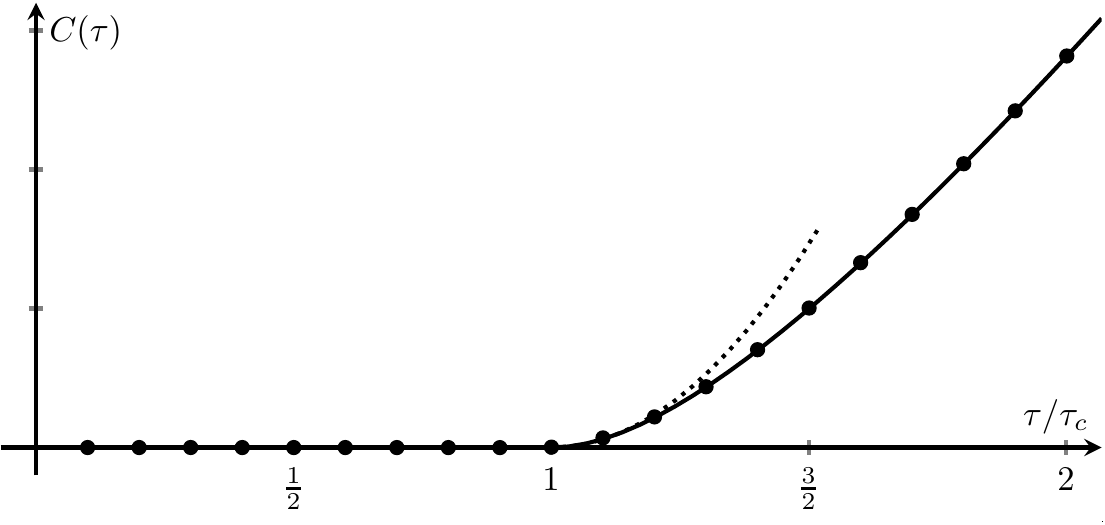}
 \caption{The figure shows the complexity as a function of the relaxation time with the parameter choice $f=1/4$, $\mu_E=3$, $\mu_I=-1$, $\sigma_E=2$, and $\sigma_I=1/2$. The solid curve shows~\eqref{complexity} while the dotted curve shows the approximation~\eqref{complexity-final} valid for $\tau\approx\tau_c$. The bullet points shows numerical data generated using~\eqref{complexity-matrix} with matrix dimension $n=1\,000$ and a mean obtained as an average over $1\,000$ realisations. It is seen that the complexity is zero up to the threshold $\tau_c$ after which it starts to grow. Numerical and analytical results are in excellent agreement.}
 \label{fig:complexity}
\end{figure}

Since our spectral measure is rotational invariant in the complex plane, we can use that for $a\in\mathbb R$, $b>0$
\begin{equation}
\int_0^{2\pi}d\theta\log\abs{a-be^{i\theta}}=2\pi\log(\min(a,b))
\end{equation}
to see that 
\begin{equation}\label{complexity}
C(\tau)\approx2\pi\int_{1/\tau}^\infty dr\rho_\text{RA}(r)\,r \log(\tau r).
\end{equation}
In order to evaluate the integral~\eqref{complexity}, we first recall that the Rajan--Abbott density~\eqref{rho-RA} only has support on a disk of radius $R=\sigma_\text{eff}\sqrt n$. 
It follows that $C(\tau)=0$ for $\tau\sigma_\text{eff}<n^{-1/2}$.
On the other the hand, the integral~\eqref{complexity} is positive for $\tau\sigma_\text{eff}>n^{-1/2}$, i.e. $C(\tau)>0$. Below, we will see that there is phase transition at $\tau_c=1/\sigma_\text{eff}\sqrt n$, see Figure~\ref{fig:complexity}, and explore the complexity~\eqref{complexity} in the `complex' region of phase space.  

To get a better intuition of what happens to the complexity~\eqref{complexity} for $\tau\sigma_\text{eff}>n^{-1/2}$, let us expand the Rajan--Abbott density~\eqref{rho-RA} in the `complex' regime in powers of $g$, we have
\begin{equation}\label{rho-RA-expand}
\rho_\text{RA}(r)=\frac{1}{\pi n\sigma_I^2}
\bigg(1-\frac g2 \sum_{k=0}^\infty a_{k,f}\Big(g \frac{r^2}{n\sigma_I^2}\Big)^{k}\bigg).
\end{equation}
with coefficients $a_{0,f}=2f,a_{1,f}=4f(1-f),a_{2,f}=6f(1-3f+2f^2),$ etc. (the explicit expressions for the coefficients are not really important to us as we will redo the sum later on). 
Using this expansion in the formula for the complexity~\eqref{complexity} gives
\begin{equation}
C(\tau)\approx\frac{2}{n \sigma_I^2} \!\!\!\int\limits_{\frac1\tau}^{\sigma_\text{eff}\sqrt{n}}\!\!\! dr  
\bigg(1-\frac g2 \sum_{k=0}^\infty a_{k,f}\Big(g \frac{r^2}{n\sigma_I^2}\Big)^{k}\bigg)\,r \log(\tau r).
\end{equation}
Here and below we have assumed that $\tau>1/\sigma_\text{eff}\sqrt n$, since (as already discussed) the complexity~\eqref{complexity} is zero for $\tau<1/\sigma_\text{eff}\sqrt n$.
Now, making a change of variables $\hat r=r/\sigma_\text{eff}\sqrt n$, we get 
\begin{equation}
C(\tau)\approx2\frac{\sigma_\text{eff}^2}{\sigma_I^2} \!\int\limits_{\frac{\tau_c}{\tau}}^{1} \! d\hat r  
\bigg(1-\frac g2 \sum_{k=0}^\infty a_{k,f}\Big(g r^2\frac{\sigma_\text{eff}^2}{\sigma_I^2}\Big)^{k}\bigg)\,
\hat r \log\Big(\frac{\tau\hat r}{\tau_c}\Big),
\end{equation}
where for notational simplicity we have used $\tau_c=1/\sigma_\text{eff}\sqrt n$. In this expression, we may integrate term by term using
\begin{multline}\label{integral}
\int\limits_{\frac{\tau_c}{\tau}}^{1} d\hat r\, \hat r^{2k+1}\log\Big(\frac{\tau\hat r}{\tau_c}\Big)
=\\
\frac{(\tau_c/\tau)^{2k+2}+(2k+2)\log(\tau/\tau_c)-1}{(2k+2)^2}.
\end{multline}
We recall that our formula for the complexity~\eqref{complexity} is only valid in the vicinity of the critical value $\tau_c$. Thus, we introduce the dimensionless parameter 
\begin{equation}
\hat\tau=\frac{\tau-\tau_c}{\tau_c}.
\end{equation}
For $1\gg\hat\tau>0$, the right-hand side in~\eqref{integral} simplifies to
\begin{equation}
\frac{(\tau_c/\tau)^{2k+2}+(2k+2)\log(\tau/\tau_c)-1}{(2k+2)^2}
=\frac{\hat\tau^2}2+O(\hat\tau^3),
\end{equation}
where the leading order term is independent of $k$. This allow us to write the complexity as
\begin{equation}
C(\tau)\approx\frac{\sigma_\text{eff}^2}{\sigma_I^2}  
\bigg(1-\frac g2 \sum_{k=0}^\infty a_{k,f}\Big(g \frac{\sigma_\text{eff}^2}{\sigma_I^2}\Big)^{k}\bigg)\,
{\hat\tau^2}+O(\hat\tau^3).
\end{equation}
By comparison with~\eqref{rho-RA-expand}, we see that
\begin{equation}\label{complexity-final}
C(\tau)\approx
\begin{cases}
\pi n\sigma_\text{eff}^2\,\rho_\text{RA}(\sigma_\text{eff}\sqrt{n})\hat\tau^2 & \hat\tau>0\\
0 & \hat\tau<0.
\end{cases}
\end{equation}
At first sight it might appear as though the complexity is proportional to $n$ for $\hat\tau>0$, but this is merely an artefact of our normalisation. In our present normalisation the Rajan--Abbott density~\eqref{rho-RA} as support on a disk with radius $\sigma_\text{eff}\sqrt{n}$ and the density is therefore inversely proportional to $\sigma_\text{eff}^2n$. The normalisation in which the connectivity matrix $J=(J_{ij})$ has support on the unit disk corresponds to
\begin{equation}
\widehat\rho_\text{RA}(z)=n\sigma_\text{eff}^2\rho_\text{RA}(z\sigma_\text{eff}\,\sqrt n)
\end{equation}
In this normalisation, the complexity reads
\begin{equation}\label{complexity-final-2}
C(\tau)\approx
\begin{cases}
\pi \widehat\rho_\text{RA}(1)\hat\tau^2 & \hat\tau>0\\
0 & \hat\tau<0.
\end{cases}
\end{equation}
For a connectivity matrix with i.i.d. entries (i.e. $f\to0$ or $f\to1$), we have $\widehat\rho_\text{RA}(1)=1/\pi$ which is the result presented by Wainrib and Touboul~\cite{Wainrib2013}. In the intermediate region ($0<f<1$) the dependence on $\hat\tau$ is the same, but the overall prefactor changes. The prefactor is given by the global spectral density of the connectivity matrix evaluated at the edge, which should not be too surprising. It is important to note that this `edge value' depends strongly on the network structure, i.e. the variances $\sigma_E$ and $\sigma_I$ as well as the fraction $f$.
We recall that asymptotically the complexity is related to the average number of equilibria by 
\begin{equation}\label{Neq=expnC}
\E\{N_\text{eq}\}\sim e^{n C(\tau)},
\end{equation}
so the changed prefactor in the complexity corresponds to an exponential change to the average number of equilibria.
When interpreting~\eqref{Neq=expnC}, it should be noted that our approximation scheme only gives the leading order contribution in the system size $n$. Thus, it does not exclude error terms which grow or remain constant for large $n$. Next-to-leading order contributions have been computed in some closely related systems~\cite{Fyodorov2016,Ipsen2018}.

\section{Outliers and network imbalance}
\label{sec:outlier}

The approximation scheme used in Section~\ref{sec:complexity} assumed that the connectivity matrix $J=(J_{ij})$ had no spectral outliers for large neuron populations $n\to\infty$. There are two main ways this assumption may be invalidated in the given framework. In Section~\ref{sec:dale}, we have seen in Figure \ref{fig:scatter} that not imposing the zero-sum constraint~\eqref{zero-sum} resulted in a number of spectral outliers scattered around the circular region which contained the majority of the eigenvalues. Breaking the network balance condition~\eqref{balance} also results in a spectral outlier; we will discuss this phenomenon towards the end of this section. 

First, we will discuss the effect on the complexity of not enforcing the zero-sum constraint~\eqref{zero-sum}. We note that the spectral outliers appearing in the absence of the zero-sum constraint (see left panel of Figure~\ref{fig:scatter}) constitute a lower-order effect in the sense that the fraction of spectral outliers (compared to the total number of eigenvalues) tends to zero as the neuron population tends to infinity ($n\to\infty$). Although not fully mathematically justified, we believe that the effect on the complexity will also be a lower-order effect, so that there is still a `complex' region (where the number of equilibria grows exponentially fast with $n$) and a `simple' region (where the number of equilibria grows less than exponentially fast with $n$). The transition between these two regions appears at the same critical value $\tau_c$. 

\begin{figure}
\centering
\includegraphics[width=.99\columnwidth]{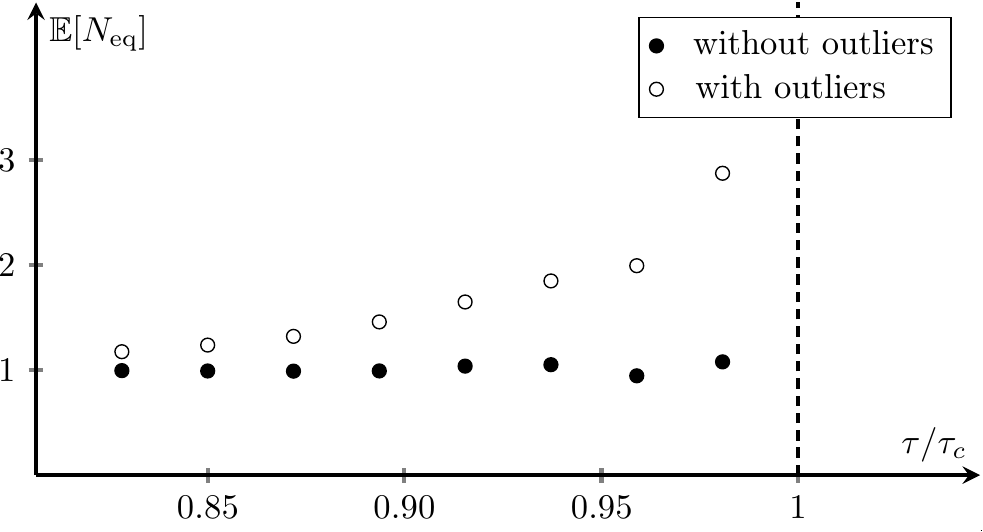}
\caption{Data point are generated using the random matrix approximation~\eqref{KacRice-approx} using parameters $n=1\,000$, $f=1/4$, $\mu_E=3$, $\mu_I=-1$, $\sigma_E=4$, and $\sigma_I=1$; the average is performed using $5\,000$ realisations. Filled data points `$\bullet$' are made using connectivity matrix~\eqref{J-no_outlier} (i.e. without outliers), while unfilled data points `$\circ$' are made using connectivity matrix~\eqref{J-outlier} (i.e. with outliers). The location of the transition to the `complex' region is indicated with a dashed line.}
\label{fig:complexity-correction}
\end{figure}

The main difference between the scenarios arising from a connectivity matrix with or without the zero-sum constraint are present in the `simple' region; this is due to the fact that the complexity is zero in this region or, in other words, the leading-order contribution vanishes in this region resulting in the lower-order contribution arising from the outliers becoming dominant.  Without outliers there will be one (and only one) equilibrium up until the critical value $\tau_c$ after which there will be an exponential explosion in the number of equilibria. With outliers there will still be only one equilibrium for short relaxation times, $\tau\ll\tau_c$, but we expect that the average number of equilibria start to grow as we approach the critical value $\tau_c$, since the spectral density is non-zero outside the disk of radius $\sigma_\text{eff}\sqrt n$ (the growth must of course be slower than exponentially fast in $n$, since the complexity is zero in the simple region). Figure~\ref{fig:complexity-correction} illlustrates this using the random matrix approximation~\eqref{KacRice-approx}.
In summary, for the system~\eqref{dyn-sys} with connectivity matrix~\eqref{J-no_outlier} we expect to find one (and only one) equilibrium below the critical value $\tau_c$, while with connectivity matrix~\eqref{J-outlier} we expect to find few (but possibly more than one) equilibria in the same region. In the later case, the true number of equilibria would be highly dependent on the realisation of the connectivity matrix $J$. The possibility of multiple equilibria can, of course, have important consequences for the dynamics.

We emphasise that the result for the case including spectral outliers should only be trusted qualitatively but not quantitatively. The issue is that the approximation which leads to~\eqref{KacRice-approx} is only fully justified in the neighbourhood of the first fixed point bifurcation. In Section~\ref{sec:complexity} we exploited that in absence of spectral outliers, this neighbourhood coincides with the neighbourhood of the critical value $\tau_c$. In the presence of spectral outliers, the first bifurcation will typically happen at some smaller value of $\tau$.

\begin{figure}
 \centering
 \includegraphics[width=.99\columnwidth]{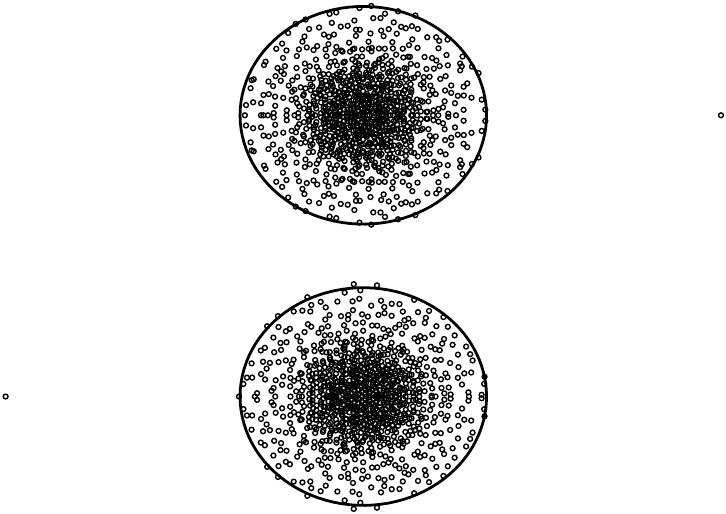}
 \caption{The figure shows the scatter plots in the complex plane for eigenvalues of connectivity 
matrix~\eqref{J-no_outlier}. Both scatter plots are generated from random matrices with $n=1\,000$, $f=1/4$, $\mu_E=3$, $\mu_I=-1$ and $\sigma_E=2$. The top plot has $\sigma_I=-13/15$ (corresponding to $\beta=+1/10$) and the bottom plot has $\sigma_I=-17/15$ (corresponding to $\beta=-1/10$). Solid curves shown are circles with radius given by~\eqref{radius}, i.e.  $R\approx34.46$. The outliers are located at $\approx\beta n=\pm100\approx\pm3R$, i.e to the right on the top plot and to the left on the bottom plot.}
\label{fig:scatter-out}
\end{figure}

Let us now turn to the effect of breaking the balance condition~\eqref{balance}. 
In abnormal brain states that are associated with pathological brain dynamics such as epilepsy, a network imbalance causes hyper-excitable unstable electrical behaviour called seizures, which are observed as pathological (spike-wave) oscillations. For this reason, network imbalance is an important scenario. We include such imbalance by replacing the balance condition~\eqref{balance} by
\begin{equation}\label{broken-balance}
\mu_Ef+\mu_I(1-f)=\beta.
\end{equation}
Here $\beta$ is a measure of imbalance, such that if $\beta>0$ ($\beta<0$) then the excitatory (inhibitory) neurons dominate the neuron population; $\beta=0$ restores the balance condition~\eqref{balance}.  
Let us first examine the eigenvalues of $M=uv^T$. The charateristic equation reads
\begin{equation}
0=\det(\lambda\one-uv^T)=\lambda^{n-1}(\lambda-v^Tu)=\lambda^{n-1}(\lambda-\beta n),
\end{equation}
hence $M$ has $n-1$ eigenvalues equal to zero and the remaining eigenvalue is equal to $\beta n$. It is known that the random connectivity matrix $J$ will inherit this spectral outlier in the following sense: the global spectrum of $J$ will still be the Rajan--Abbott density~\eqref{rho-RA} on the disk of radius $\sigma_\text{eff}\sqrt n$, but there will also be an eigenvalue in the neighbourhood of $\beta n$ (this is illustrated on figure~\ref{fig:scatter-out}); we will not prove the details here but refer the reader to~\cite{tao2013outliers} for a mathematical description of this phenomenon.
The number of equilibria should be affected by the eigenvalue with the largest real part (i.e. the rightmost eigenvalue), so we would expect an excitatory dominated neuron imbalance ($\beta>0$) having essential effects, but that a inhibitory dominated neuron imbalance ($\beta<0$) would inconsequential. It important to note the location of the outlier grows as linearly with neuron population $n$ (we assume that $\beta$ is of order unity), while the radius of the disk containing the global spectrum only grows as $\sqrt n$. Thus, the effect of breaking the network balance happens on a completely different scale than all other effects discussed in this paper and even a small imbalance may imply important effects.

\section{Discussion and Outlook}
\label{sec:discuss}


In this paper we have combined two important results in the theory of random neural networks to show how a more anatomically realistic network connectivity in the form of Dale's Law impacts on the transition to a `complex' phase. The transition is affected by the differences in variances between excitatory and inhibitory neurons ($\sigma_E \neq \sigma_I$) and the proportion of each population $f$.  Changes to these variables are reflected in changes to the effective variance $\sigma_{\mathrm{eff}}=(f\sigma_E^2+(1-f)\sigma_I^2)^{1/2}$. A transition takes place at the critical value $\tau_c=(\sigma_{\mathrm{eff}} \sqrt{n})^{-1}$. For the case of a uniformly distributed connectivity matrix that has i.i.d. entries with zero mean and variance $\sigma$, this becomes $\tau_c=(\sigma\sqrt{n})^{-1}$, as originally established by~\cite{Sompolinsky1988}. It is worth noting that a stability-instability phase transition in a linearised model was found at the same critical value in the seventies~\cite{May1972}; in this latter context it is often referred to as the May--Wigner transition.

In neural networks where connections are chosen as i.i.d. random variables with zero mean and variance $\sigma$ the eigenvalue spectrum of the connectivity matrix is uniform on a disk. However, for the case of a more anatomically realistic connectivity where Dale's law is implemented into the structure of the connectivity matrix, the eigenvalue spectrum is no longer uniform and one of the results of this paper was to significantly simplify the expression for the non-uniform spectral density derived in~\cite{Rajan2006}.  In our expression it is very intuitive to see how different population proportions and variances affect the density as well as how the expression collapses back into a uniform density when the population variances are equivalent.   
Note that if the variances are not different, then changing the proportion $f$ has no effect and the spectral density of the eigenvalues remains uniform.  

The expression for the non-uniform spectral density was then used in combination with a random matrix approximation of the multivariate Kac--Rice formula~\cite{Wainrib2013}. Incorporating Dale's Law (without spectral outliers) we were able to establish that a phase transition occurs by computing the number of equilibria (fixed points) or `Complexity' of the system. This transition divides phase space into two regions: a `simple' phase with a single equilibrium and a `complex' phase where the number of equilibria grows exponentially with the neuron population. 
Dynamics in the `simple' phase are expected to be stable meaning that all trajectories converge towards the only equilibrium of the system, while dynamics in the `complex' phase is expected to be chaotic. As a na\"ive mental picture, we might imagine that most trajectories in the `complex' phase are bouncing chaotically around between a large number of unstable equilibria.

The effects on the complexity of incorporating different connectivity structures via Dale's law on the phase transition is encapsulated by Figure \ref{fig:complexity}.  This figure shows how the complexity changes with the ratio $\tau/\tau_c$.  Both the transition point $\tau_c$ and rate at which the complexity increases beyond this transition point are affected by the variances $\sigma_E, \sigma_I$ and population proportions $f$. However, the critical exponent of the control parameter $\hat\tau=(\tau-\tau_c)/\tau_c$ is unaffected and equal to $2$, which is consistent with the common lore regarding phase transitions in statistical systems. Under more general network connectivity structures we believe that there will still be a transition between a `simple' and a `complex' phase characterised by having few or many equilibria, respectively (`many' means growing exponentially fast with the neuron population, and `few' means growing less than exponentially fast). We believe that there is a universal scaling limit connected to this transition and, in particularly, that the critical exponent $2$ is universal. Away from the critical region the number of equilibria  is still important for the dynamic of the system, but we do not expect this behaviour to be universal. Neither do we expect the location of the transition to be universal. Here, we understand `universality' as properties being unaffected by the finer details of the network structure.  Thus, understanding patient specific network structures are fundamental to define when pathological transitions could occur.  Alternatively, we could understand `universality' as properties being unaffected by the finer details of the probability weights chosen for the connectivity given a certain network structure. In this latter interpretation, the complexity might be universal, but of less practical value.

In this work we also discuss eigenvalue outliers (i.e. eigenvalues outside the disk with radius $\sigma_\text{eff}\sqrt n$) that are generated by two sources. The first source is from implementing Dale's law into the connectivity matrix, and can be eliminated via an input-output condition using a projection operator (as illustrated by Figure~\ref{fig:scatter}). The outliers are a lower order effect of the eigenvalue spectrum (the fraction of outliers tends to zero as the neuron population tends to infinity), and their contribution to the complexity is believed to be lower order as well. Such lower contribution can nonetheless have fundamental importance for the dynamical properties of the system. In Section~\ref{sec:outlier}, we argued that spectral outliers would give rise to multiple equilibria in the `simple' phase rather than just one equilibrium as were found in the absence of outliers. It should be clear that going from one to multiple equilibria may fundamentally change dynamical properties of the system, e.g. we might imagine trajectories bouncing between a few unstable equilibria. In fact, a distinction between having one or multiple equilibria might be as important as the distinction between have few and many.

The second source of eigenvalue outliers comes from breaking the balance condition by introducing an `imbalance' parameter $\beta$. In this scenario, there is an outlier in the neighbourhood of $\beta n$. The important observation to be made here is that the location of this outlier scales with $n$ while the location of all other eigenvalues (including outliers arising from introducing Dale's Law without imposing the input-output condition) scales with $\sqrt n$. Thus, the effect of breaking the network balance happens on a completely different scale than all other effects discussed in this paper, and even a small imbalance can have significant consequences on the system dynamics. It is often assumed for pathological behaviour such as those found in epileptic states, that the connectivity matrix is excitatory dominated ($\beta>0$).  In this case, the outlier is the largest (rightmost) eigenvalue, and the location of this eigenvalue determines the onset of the increase in the number equilibria, which effects the stability of the system.  Hence, a network imbalance condition such as hyper-excitability, can determine the change in the system stability more than the network structure and nonlinear coupling function.

It is important to note that neural model used in this research is abstracted and does not have a direct correspondence to physiology unlike biophysical spiking models.  It is a first order rate-based neural model with a discretised spatial field, and although it can be derived from networked neural mass models, it does so at the cost of sacrificing synaptic dynamics, which are fundamental to brain dynamics.  It would be interesting to expand this model to a second order one to give it synaptic dynamics, either current-based or conductance-based and examine the role of these in combination with different network structures on stability transitions.  This model also assumes a single time-constant, where as a distribution of values would be more realistic as well as different parameterisations of the coupling function.  Work using dynamical mean-field theory has shown that results of this model can be projected onto more physiologically realistic spiking models such as leaky integrate-and-fire models~\cite{Ostojic2014}, which normally cannot be analysed in this way. It would be interesting to see if the results presented here also hold for these more realistic neural models. Further, it is currently an open problem to compute the actual number of fixed points through a large scale simulation, it would be good to conclusively verify these results without numerical approximations. 

From a brain dynamics perspective, this paper shows that when more anatomically realistic network connectivities are used that have heterogeneous structures, they have a non-trivial effect on the phase transition of random neural networks. These connectivities are statistically characterised with block structures that have different sizes, means and variances.  
In the brain, these connectivity parameters are regulated by many physiological and anatomical processes and it has been experimentally and clinically observed that certain network structures are more susceptible to a transition into instability~\cite{Khambhati2015}.  This becomes especially important as the brain is often considered to be operating in a critical state i.e. close to a transition point~\cite{Fontenele2019}.  If the brain does indeed operate in such a critical regime, then the network connectivity plays a crucial role in brain-state transitions.  The study of brain diseases such as epilepsy are usually conceptualised as a brain state transition into a pathological state such as a seizure, which is both hyper-excitable and hyper-synchronised.  However, the main problem here is that such transitions and hence diseases are very patient-specific depending on the patients individual brain connectivity.  This research illustrates that a large class of network connectivities that can be identified statistically, are more susceptible to such transitions.  This could be a way of classifying certain patient-specific network structures in order to quantify the susceptibility to a transition to an unstable pathological state such as a seizure.  This susceptibility is often defined as excitability in epileptogenesis, a clinical term that describes the underlying process by which a brain develops seizures, which is currently poorly understood.  A quantitative description of this susceptibility based on an individuals brain network structure from neuroimaging data could be used to tailor diagnosis and treatment, which is one of the primary goals of 21st century precision medicine.

\paragraph*{Acknowledgement}

We acknowledge financial support by ARC Centre of Excellence for Mathematical and
Statistical frontiers (JRI) and Sir John Eccles Research Fellowship (ADHP).

\end{document}